\documentclass[twocolumn]{aastex63}
\usepackage{threeparttable}
\usepackage{bm}
\usepackage{chngpage}
\usepackage{graphicx}
\usepackage{mathrsfs}
\usepackage{amsmath}

\def\apjl{ApJL}

\shorttitle{A Magnetar-Asteroid Impact Model for an FRB/XRB}
\shortauthors{Dai}

\begin{document}

\title{A Magnetar-Asteroid Impact Model for FRB 200428 Associated with an X-ray Burst \\ from SGR 1935+2154}

\author[0000-0002-7835-8585]{Z. G. Dai}
\affil{School of Astronomy and Space Science, Nanjing University, Nanjing 210023, China; dzg@nju.edu.cn}
\affil{Key Laboratory of Modern Astronomy and Astrophysics (Nanjing University), Ministry of Education, Nanjing, China}

\begin{abstract}
Very recently, an extremely bright fast radio burst (FRB) 200428 with two sub-millisecond pulses was discovered to come from the direction of the Galactic magnetar SGR 1935+2154, and an X-ray burst (XRB) counterpart was detected simultaneously. These observations favor magnetar-based interior-driven models. In this Letter, we propose a different model for FRB 200428 associated with an XRB from SGR 1935+2154, in which a magnetar with high proper velocity encounters an asteroid of mass $\sim10^{20}\,$g. This infalling asteroid in the stellar gravitational field is first possibly disrupted tidally into a great number of fragments at radius $\sim {\rm a\,\,few}$ times $10^{10}\,$cm, and then slowed around the Alfv$\acute{\rm e}$n radius by an ultra-strong magnetic field and in the meantime two major fragments of mass $\sim 10^{17}\,$g that cross magnetic field lines produce two pulses of FRB 200428. The whole asteroid is eventually accreted onto the poles along magnetic field lines, impacting the stellar surface, creating a photon-e$^\pm$ pair fireball trapped initially in the stellar magnetosphere, and further leading to an XRB. We show that this gravitationally-powered model can interpret all of the observed features self-consistently.
\end{abstract}

\section{Introduction}\label{intr}
Fast radio bursts (FRBs) are mysterious millisecond-duration transients of GHz radio emission \citep{Lorimer2007,Thornton2013} because their physical origin and mechanism remain unknown \citep[for observational and theroretical reviews see][]{Petroff2019,Cordes2019,Katz2019,Platts2019}. This year, the first light of understanding FRBs seems to appear due to two discoveries. First, a $\sim16\,$day-period repeating source FRB 180916.J0158+65 was discovered \citep{CHIME2020a}. An activity of a longer period $\sim160\,$days for the first repeating FRB 121102 was then reported \citep{Rajwade2020}. These observations suggest that FRBs could arise from periodic objects such as precessing magnetars \citep{Yang2020,Levin2020,Zanazzi2020} or magnetized neutron stars in binaries \citep{Dai2020,Lyutikov2020,Ioka2020}. For the former models, however, starquake-like events occurring at a stellar fixed region are required to produce an FRB 180916.J0158+65-like periodic phenomenon.

Second, an extremely bright FRB 200428 with two pulses of intrinsic durations $\sim0.60$\,ms and $0.34$\,ms from the direction of the Galactic magnetar SGR 1935+2154 was reported \citep{chime20b,boch20}. The two pulses are separated by $\sim28.9$\,ms. This burst was detected to have an average fluence of $700$\,kJy\,ms and $1.5\pm0.3\,$MJy\,ms by the CHIME and STARE2 telescopes, respectively, which imply the isotropic-equivalent energy release of $E_{\rm CHIME}=3^{+3.0}_{-1.6}\times 10^{34}$\,erg and $E_{\rm STARE2}=(2.2\pm0.4)\times 10^{35}$\,erg in two different frequency bands for the source's distance $D\sim10$\,kpc. Very fortunately, an X-ray burst (XRB) with two corresponding pulses associated with FRB 200428 was simultaneously detected by high-energy satellites such as {\em Insight}-HXMT \citep{Li2020}, AGILE \citep{tav20}, INTEGRAL \citep{mere20}, and Konus-Wind \citep{rid20}. The isotropic-equivalent emission energy release of the XRB in the soft X-ray to soft gamma-ray energy band is $E_{\rm X}\sim (0.8-1.2)\times 10^{40}(D/10\,{\rm kpc})^2\,$erg \citep{Li2020,mere20,rid20,tav20}. The spectrum seems to be fitted by either a cutoff power law model \citep{Li2020,mere20,rid20,tav20} or a double-temperature blackbody model \citep[$kT_1\sim11\,$keV and $kT_2\sim 30\,$keV,][]{mere20,rid20}.

The physical parameters of the magnetar SGR 1935+2154 include the rotation period $P\simeq3.24$\,s, spin-down rate ${\dot{P}}\simeq 1.43\times 10^{-11}\,{\rm s}\,{\rm s}^{-1}$, surface dipole magnetic field strength $B_{\rm s}\simeq2.2\times10^{14}~{\rm G}$, and spin-down age $t\sim3.6$ kyr \citep{isr16}. The source is hosted in the Galactic supernova remnant (SNR) G57.2+0.8 \citep{gae14}. However, some estimates of the distance $D$ and age of the SNR remain highly debated, e.g., $D$ is in a range of $\sim 6.6$ to $\sim 12.5\,$kpc \citep{pav13,surn16,koth18,zhou20,Zhong2020}.  Inferred recently from the observed dispersion measure and Faraday rotation measure, $D$ turns out to be $9.0\pm2.5$\,kpc \citep{Zhong2020}. Although this range implies that the isotropic-equivalent energy release of FRB 200428 is close to the low energy end of cosmological FRBs \citep{chime20b,boch20}, the association of the FRB with SGR 1935+2154 clearly indicates a magnetar origin at least for some FRBs. Based on the frame of a magnetar, some models for the association of an FRB/XRB were discussed \citep{Lyut2020,Margalit2020,Lu2020}, in which both FRBs and XRBs are triggered by starquake-like explosions and powered magnetically. We call these models interior-driven ones.

\begin{figure}
\hspace{-17.5mm}
\includegraphics[width=0.64\textwidth, angle=0]{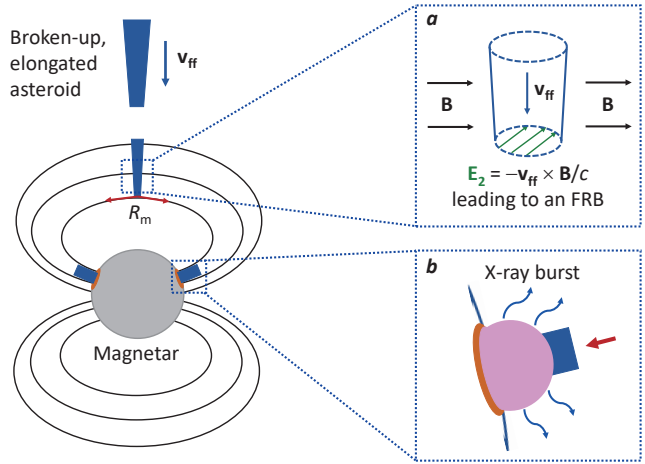}
\caption{Schematic picture of magnetar-asteroid impact. A rocky asteroid of mass $m_{\rm a}\sim 10^{20}\,$g is disrupted tidally into a great number of fragments at $R_{\rm d}\sim {\rm a\,\,few}\times 10^{10}\,$cm, of which two major fragments of mass $m\sim 10^{17}\,$g are then distorted tidally at breakup radius $R_{\rm b}\sim 10^9\,$cm. A broken-up, elongated fragment (blue shaded region) freely falls in the gravitational field of a magnetar (gray shaded region) below $R_{\rm b}$ and meanwhile crosses magnetic field lines downwards to the magnetic interaction radius $R_{\rm m}$, and then is accreted onto the poles along magnetic field lines. {\em Panel a}: An electric field (${\bf E_2}$) induced outside of the fragment by this crossing has such a large component parallel to the magnetic field around $R_{\rm m}$ that electrons are torn off the asteroidal surface and accelerated to ultra-relativistic energies instantaneously. The electrons subsequently move along magnetic field lines and their coherent curvature radiation causes an FRB. {\em Panel b}: The asteroid eventually collides with the stellar surface, creating a hot spot (orange shaded region) and an e$^\pm$-pair fireball (pink shaded region). The fireball is initially trapped by closed magnetic field lines and subsequently expands relativistically. Its energy may be released via the following processes: X-rays are emitted from a photosphere and superposition of thermal emission from different photospheres (corresponding to different fragments impacting the stellar surface) and nonthermal emission from collisions between some shells in the relativistic fireball could generate an XRB (blue wave arrows).}
\label{fig1}
\end{figure}

In this Letter, we propose a different model for the association of FRB 200428 with an XRB from SGR 1935+2154, in which a magnetar encounters an asteroid. We show that such an impact can interpret all of the observed features self-consistently. The impact and radiation physics were discussed in detail when a {\em moderately magnetized pulsar} encounters an asteroid \citep{Dai2016}, in which case an asteroid can freely fall onto the stellar surface and lead to a bright cosmological FRB. For a magnetar, however, an asteroid during its free infall must be impeded around the Alfv$\acute{\rm e}$n radius by an ultra-strong magnetic field and then accreted onto the poles along magnetic field lines, colliding with the stellar surface instantaneously and generating an XRB (see Figure \ref{fig1}). Although it is undetected at cosmological distances, such an XRB in the Galaxy is bright enough to be observed by X-ray satellites \citep[for a discussion see][]{Dai2016}. It should be pointed out that this gravitationally-powered model does not exclude magnetar-based interior-driven models \citep[for a brief summary on four kinds of energy source see][]{Dai2017}, some of which, together with our mechanism, might be able to take place for an FRB/XRB. This Letter is organized as follows. We describe our model in Section 2 and constrain the model parameters in Section 3. We present our conclusions in Section 4.

\section{The Model}\label{model}

The Hubble Space Telescope observations of the magnetar SGR 1935+2154 with an ultra-strong magnetic dipole field show that this magnetar is moving at a high proper velocity $V_{\rm p}=(600\pm400)(D/10\,{\rm kpc})\,{\rm km}\,{\rm s}^{-1}$ \citep{Levan2018}. This leads to our assumption that the magnetar could catch up with decelerated supernova ejecta and encounter a rocky asteroid of mass $m_{\rm a}\sim10^{20}\,$g. Such an asteroid, which is further assumed to include an iron-nickel component of mass $m\sim 10^{17}\,$g, could be one of the asteroids hosted by the magnetar's progenitor before supernova explosion \citep[corresponding to the first scenario of][]{Tremaine1986} or could be formed during the collapse of some parts of supernova ejecta in SNR G57.2+0.8 or could happen to wander nearby the magnetar from the outside. The rate of asteroid-neutron star collisions in the Milky Way has been estimated in different scenarios \citep{Tremaine1986,Wasserman1994,Siraj2019}. In particular, the rate of such events may be so high as $\sim 0.1-1$ per day under reasonable conditions \citep{Wasserman1994} and even could reach $\sim 10$ per day at flux $\sim1\,$Jy in radio band \citep{Siraj2019}.

The stellar mass, radius, and surface dipole field strength are taken to be $M$, $R_*$, and $B_{\rm s}$, respectively. The asteroid is first disrupted tidally into a great number of fragments in the stellar gravitational field at radius $R_{\rm d}\sim (M/m_{\rm a})^{1/3}r_{\rm a}\sim6.1\times10^{10}(M/1.4M_\odot)^{1/3}\,$cm, where $r_{\rm a}\sim2.0\times 10^6(m_{\rm a}/10^{20}\,{\rm g})^{1/3}\,$cm is the rocky asteroid's original radius, and then two major fragments of mass $\sim m$ are further distorted tidally by the magnetar at breakup radius, $R_{\rm b} = 1.3\times 10^9(m/10^{17}\,{\rm g})^{2/9}(M/1.4M_\odot)^{1/3}\,{\rm cm}$, where the fragmental tensile strength and original mass density have been taken for iron-nickel matter \citep{Colgate1981}. The time difference of arrival of leading and lagging parts of a fragment at any radius is estimated by
\begin{eqnarray}
\Delta t & \simeq & \frac{12r_0}{5}\left(\frac{R_{\rm b}}{GM}\right)^{1/2}\nonumber\\ & = & 0.57\left(\frac{m}{10^{17}\,{\rm g}}\right)^{4/9}\left(\frac{M}{1.4M_\odot}\right)^{-1/3}\,{\rm ms},\label{Dt}
\end{eqnarray}
where $r_0$ is the fragmental original radius \citep{Dai2016}. This timescale is independent of free-fall radius ($R$) and thus can be considered as the duration of an FRB \citep{Geng2015}. A requirement of the first-pulse intrinsic duration $\Delta t\sim 0.6\,$ms of FRB 200428 \citep{chime20b,boch20} leads to the fragmental mass
\begin{equation}
m\simeq 1.1\times 10^{17}\left(\frac{\Delta t}{0.6\,{\rm ms}}\right)^{9/4}\left(\frac{M}{1.4M_\odot}\right)^{3/4}\,{\rm g}.\label{mass}
\end{equation}
In the following, we discuss the geometry of an FRB-emitting region and observed features of an FRB/XRB.

\subsection{Geometry of an FRB-Emitting Region}
\begin{figure}
\vspace{-3mm}
\hspace{-25mm}
\includegraphics[width=0.73\textwidth, angle=0]{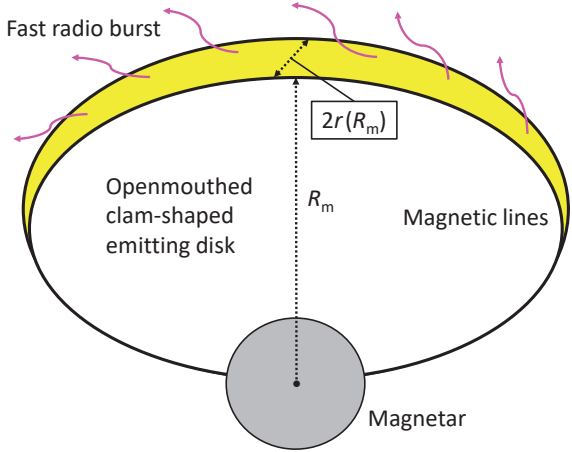}
\caption{Schematic picture of the geometry of an FRB-emitting region for relativistic electrons (with bulk Lorentz factor $\gamma$) moving from right to left along magnetic field lines of a magnetar (gray shaded region). The emitting disk looks like an openmouthed clam, on which mouth (yellow shaded region) relativistic electrons radiate an FRB (red wave arrows). The inclination angle from the symmetric plane of the emitting disk is approximated by $\theta_{\rm i}\simeq r(R_{\rm m})/R_{\rm m}\ll1/\gamma$ for FRB 200428 and thus the FRB's solid angle is given by $\Delta\Omega\simeq2\pi\times \max(\theta_{\rm i},1/\gamma)=2\pi/\gamma$.}
\label{fig2}
\end{figure}

\cite{Dai2016} analyzed the fragmental size and mass density as functions of $R$ during the free-fall. Physically, the fragment is initially elongated as an incompressible flow from $R_{\rm b}$ and subsequently further transversely compressed to a cylinder \citep{Colgate1981}. Here we present two evolutional results. First, the radius of the cylindrical fragment at $R$ is written as
\begin{eqnarray}
r = 1.9\times 10^4\left(\frac{\Delta t}{0.6\,{\rm ms}}\right)^{1/2}\left(\frac{R}{10^7\,{\rm cm}}\right)^{1/2}\,{\rm cm}.\label{radius}
\end{eqnarray}
Second, the free-fall fragment is significantly affected by the stellar magnetic field, whose interaction radius ($R_{\rm m}$) is approximately equal to the Alfv$\acute{{\rm e}}$n radius \citep{Ghosh1979}. At the latter radius, the kinetic energy density of the fragment is equal to the magnetic energy density. Assuming that the magnetic dipole moment $\mu=B_{\rm s}R_*^3$ and the free-fall velocity $v_{\rm ff}=(2GM/R)^{1/2}$, we thus derive
\begin{eqnarray}
R_{\rm m} & \simeq & 1.2\times 10^7\left(\frac{\Delta t}{0.6\,{\rm ms}}\right)^{-1/18} \left(\frac{M}{1.4M_\odot}\right)^{-15/54}\nonumber \\ & & \times \left(\frac{\mu}{2.2\times 10^{32}\,{\rm G}\,{\rm cm}^3}\right)^{4/9}\,{\rm cm},\label{Rm1}
\end{eqnarray}
where following \cite{Litwin2001} we have assumed that at $R\sim R_{\rm m}$ the plasma in the fragment becomes thoroughly ``threaded'' by the magnetic field so that a strong electric field ${\bf E_2}$ is induced and meanwhile the fragment is significantly slowed because around this radius the external magnetic field is commonly believed to penetrate the plasma \citep{Lamb1973,Burnard1983,Hameury1986}. Inserting Equation (\ref{Rm1}) into Equation (\ref{radius}), we further obtain the cylindrical radius at $R_{\rm m}$,
\begin{eqnarray}
r(R_{\rm m}) & \simeq & 2.1\times 10^4\left(\frac{\Delta t}{0.6\,{\rm ms}}\right)^{17/36} \left(\frac{M}{1.4M_\odot}\right)^{-15/108}\nonumber \\ & & \times \left(\frac{\mu}{2.2\times 10^{32}\,{\rm G}\,{\rm cm}^3}\right)^{2/9}\,{\rm cm}.\label{rRm}
\end{eqnarray}
This radius together with $R_{\rm m}$ determines an FRB-emitting region (i.e., yellow shaded region in Figure \ref{fig2}), whose relevant disk looks like an {\em openmouthed clam} and its inclination angle from the symmetric plane is
\begin{eqnarray}
\theta_{\rm i}\simeq\frac{r(R_{\rm m})}{R_{\rm m}} & \simeq & 1.7\times 10^{-3}\left(\frac{\Delta t}{0.6\,{\rm ms}}\right)^{19/36} \left(\frac{M}{1.4M_\odot}\right)^{15/108}\nonumber \\ & & \times \left(\frac{\mu}{2.2\times 10^{32}\,{\rm G}\,{\rm cm}^3}\right)^{-2/9}.\label{thetai}
\end{eqnarray}
It will be seen in Section \ref{const} that this angle is much smaller than the inverse of the typical bulk Lorentz factor of FRB-emitting electrons (i.e., $\gamma\sim 120$).

\subsection{Features of an FRB/XRB}
\subsubsection{An FRB}
When the fragment crosses the stellar magnetic field lines over $R_{\rm m}$, as shown in \cite{Dai2016}, an electric field (${\bf E}_2=-{\bf v}_{\bf ff}\times {\bf B}/c$) is not only induced outside of the fragment but it also has such a strong component parallel to the stellar magnetic field that electrons are torn off the fragmental surface and accelerated to ultra-relativistic energies instantaneously. Subsequent movement of these electrons along magnetic field lines leads to coherent curvature radiation. This emission component can account for the following features of an FRB.

First, as they move along a magnetic field line with curvature radius $\rho_c$ at radius $R_{\rm m}$, ultra-relativistic electrons produce curvature radiation and their typical Lorentz factor $\gamma$ at $R_{\rm m}$ can be calculated by
\begin{eqnarray}
\gamma & \equiv & \chi\gamma_{\rm max}\simeq\chi\left(\frac{6\pi e |E_2|}{\sigma_TB^2}\right)^{1/2}\nonumber\\ & \simeq & 140\chi\left(\frac{\Delta t}{0.6\,{\rm ms}}\right)^{-5/72}\left(\frac{M}{1.4M_\odot}\right)^{-7/72}\nonumber\\ & & \times \left(\frac{\mu}{2.2\times 10^{32}\,{\rm G}\,{\rm cm}^3}\right)^{1/18},\label{gamma}
\end{eqnarray}
where $\sigma_T$ is the Thomson scattering cross section, the parameter $\chi$ is introduced by \cite{Dai2020}, and the maximum Lorentz factor $\gamma_{\rm max}$ is given by Equation (12) of \cite{Dai2016}. Therefore, the characteristic frequency of curvature radiation observed at an angle ($\theta_{\rm v}$) from the symmetric plane of the openmouthed-clam-shaped disk becomes
\begin{eqnarray}
\nu_{\rm curv} & \simeq & 2.0\chi^3\delta\left(\frac{\Delta t}{0.6\,{\rm ms}}\right)^{-5/24}\left(\frac{M}{1.4M_\odot}\right)^{-7/24}\nonumber\\ & & \times \left(\frac{\mu}{2.2\times 10^{32}\,{\rm G}\,{\rm cm}^3}\right)^{1/6}\left(\frac{\rho_{\rm c}}{10^7\,{\rm cm}}\right)^{-1}\,{\rm GHz}\nonumber\\ & \simeq & 2.5\chi^3\delta\left(\frac{\Delta t}{0.6\,{\rm ms}}\right)^{-11/72}\left(\frac{M}{1.4M_\odot}\right)^{-1/72}\nonumber\\ & & \times \left(\frac{\mu}{2.2\times 10^{32}\,{\rm G}\,{\rm cm}^3}\right)^{-5/18}\,{\rm GHz},\label{nu}
\end{eqnarray}
where $\delta=1/\{2\gamma^2[1-\beta\cos(\theta_{\rm v}-\theta_{\rm i})]\}$ is the factor related with the Doppler effect: $\delta=1$ for $\theta_{\rm v}\le\theta_{\rm i}$, $\delta\simeq 1/[\gamma(\theta_{\rm v}-\theta_{\rm i})]^2$ for $1/\gamma\ll\theta_{\rm v}-\theta_{\rm i}\ll 1$, and otherwise $\delta\simeq 1/(2\gamma^2)$ \citep[cf.][]{Lin2020}. It is noted that the second equality of Equation (\ref{nu}) has used $\rho_{\rm c}=0.635R_{\rm m}$ near the equator from Appendix G of \cite{Yang2018}.

Second, by considering the region of coherent curvature radiation (as shown in Figure \ref{fig2}), the total luminosity ($L_{\rm tot}$) of a beamed FRB has been given by Equation (15) of \cite{Dai2016},
\begin{eqnarray}
L_{\rm tot} & \sim & 2.0\times 10^{36}\frac{1}{\chi^3}\left(\frac{m}{10^{17}\,{\rm g}}\right)^{8/9}\left(\frac{M}{1.4M_\odot}\right)^{19/12}\nonumber\\ & & \times \left(\frac{\mu}{2.2\times10^{32}\,{\rm G}\,{\rm cm}^3}\right)^{3/2}\left(\frac{R_{\rm m}}{10^7\,{\rm cm}}\right)^{-23/4}\nonumber\\ & & \times \left(\frac{\rho_{\rm c}}{10^7\,{\rm cm}}\right)^{-1}\,{\rm erg}\,{\rm s}^{-1},\label{Lr}
\end{eqnarray}
where the factor $1/\chi^3$ is added as $\chi$ is introduced in Equation (\ref{gamma}). The isotropic-equivalent emission energy of the FRB observed at angle $\theta_{\rm v}$ is thus estimated by
\begin{eqnarray}
E_{\rm radio}\simeq \frac{\delta^3}{f}\times L_{\rm tot}\times \Delta t,\label{Er}
\end{eqnarray}
where $f\equiv\Delta\Omega/(4\pi)\simeq1/(2\gamma)$ is the beaming factor because the FRB's solid angle $\Delta\Omega\simeq2\pi\times \max(\theta_{\rm i},1/\gamma)=2\pi/\gamma$ (see Figure \ref{fig2}). Inserting Equations (\ref{mass}), (\ref{Rm1}), (\ref{gamma}), and (\ref{Lr}) into Equation (\ref{Er}), therefore, we obtain
\begin{eqnarray}
E_{\rm radio} & \sim & 1.4\times 10^{35}\frac{\delta^3}{\chi^2}\left(\frac{\Delta t}{0.6\,{\rm ms}}\right)^{119/36} \left(\frac{M}{1.4M_\odot}\right)^{145/36}\nonumber \\ & & \times \left(\frac{\mu}{2.2\times 10^{32}\,{\rm G}\,{\rm cm}^3}\right)^{-13/9}\,{\rm erg}.\label{Er1}
\end{eqnarray}

\subsubsection{An XRB}\label{xrb}
If an infalling asteroid was diamagnetic, it would become an accretion sheet near a magnetar under the compression of an ultra-strong magnetic field \citep{Colgate1981}. In this case, a fan-shaped hot plasma would be ejected during the collision of the asteroid with stellar surface, even though the collision physics is a little complicated. However, following \cite{Litwin2001}, as widely argued in the literature  \citep[e.g.,][]{Lamb1973,Burnard1983,Hameury1986}, a broken-up asteroid is thoroughly threaded by an ultra-strong magnetic field so that the asteroid is significantly slowed around $R_{\rm m}$ and then moves along magnetic field lines from $R_{\rm m}$ onto the poles, possibly giving rise to two accretion columns \citep{Mes1992}. The total asteroid-magnetar gravitational energy is given by
\begin{eqnarray}
E_{\rm G} \simeq \frac{GMm_{\rm a}}{R_*} & \sim & 1.9\times 10^{40}\left(\frac{m_{\rm a}}{10^{20}\,{\rm g}}\right)\nonumber \\ & & \times  \left(\frac{M}{1.4M_\odot}\right)\left(\frac{R_*}{10^6\,{\rm cm}}\right)^{-1}\,{\rm erg}.\label{Eg}
\end{eqnarray}
This energy is released in a timescale \citep{Dai2016}
\begin{eqnarray}
t_{\rm a} & \simeq & \frac{12r_{\rm a}}{5}\left(\frac{R_{\rm d}}{GM}\right)^{1/2}\nonumber\\ & \sim & 0.1\left(\frac{m_{\rm a}}{10^{20}\,{\rm g}}\right)^{1/3}\left(\frac{M}{1.4M_\odot}\right)^{-1/3}\,{\rm s}.\label{ttot}
\end{eqnarray}
When the asteroid collides with the stellar surface, a conical hot plasma (or more likely a hemispherical fireball, as shown by the pink shaded region in Figure \ref{fig1}) is ejected, which is different from a fan-shaped outflow discussed by \cite{Colgate1981}. This hot fireball is initially trapped by closed magnetic field lines and thus its temperature is estimated by \citep{Thompson1995}
\begin{eqnarray}
T_{\rm fb}\sim\left(\frac{B^2}{8\pi a}\right)^{1/4}=1.8\times 10^{10}\left(\frac{B}{10^{14}\,{\rm G}}\right)^{1/2}\,{\rm K},\label{Tfb}
\end{eqnarray}
where $a$ is the radiation energy density constant and $B$ is the magnetic field strength. Owing to the fact that $kT_{\rm fb}$ is greater than the electron rest energy, a huge optical depth to photon-photon annihilation reaction in the fireball, $\tau_{\gamma\gamma}\sim\sigma_TE_{\rm G}/[(2\pi r_{\rm i}^2)(2.7kT_{\rm fb})]\sim 3\times 10^{10}(E_{\rm G}/10^{40}\,{\rm erg})(r_{\rm i}/10^5\,{\rm cm})^{-2}(T_{\rm fb}/10^{10}\,{\rm K})^{-1}$ (where $r_{\rm i}$ is the fireball's initial radius and $2.7kT_{\rm fb}$ is the mean thermal photon energy), inevitably leads to a dense population of e$^\pm$ pairs, further making the fireball become highly collisional and opaque. This thus drives a relativistic outflow that traps radiation near the star and releases energy at radius much larger than $R_*$ \citep{Thompson1995,Kaspi2017}. This process resembles what happens in cosmological gamma-ray bursts \citep{Kumar2015}. On one hand, thermal X-rays are emitted from a photosphere of the outflow and superposition of thermal emission from different photospheres (corresponding to different fragments colliding with the stellar surface) possibly generates a multi-temperature blackbody spectrum of an XRB, whose duration is of order $\sim t_{\rm a}$ estimated by Equation (\ref{ttot}). On the other hand, collisions between different shells in the relativistic outflow could produce an additional nonthermal emission.

\section{Constraints on Model Parameters}\label{const}

FRB 200428 has two pulses separated by $\sim28.9$\,ms. Their intrinsic durations are $\sim 0.6\,$ms and $\sim0.34$\,ms, respectively, and their fluence ratio is $\xi\sim480/220=2.2$ \citep{chime20b}. The isotropic-equivalent energy release of {\em the first pulse} as an example is thus $E_{\rm radio}\simeq [\xi/(1+\xi)]\times(E_{\rm CHIME}+E_{\rm STARE2})\sim 1.7\times 10^{35}(\Delta t/0.6\,{\rm ms})\,$erg for $D\sim10\,$kpc \citep[for $D$ also see][]{Zhong2020}, where $E_{\rm CHIME}$ and $E_{\rm STARE2}$ are the isotropic-equivalent radio emission energies observed by the CHIME and STARE2 telescopes \citep{chime20b,boch20}, respectively. Therefore, we can constrain the model parameters.

 First, from Equation (\ref{mass}), we find the fragmental mass
 \begin{equation}
 m\sim 1.1\times 10^{17}\left(\frac{\Delta t}{0.6\,{\rm ms}}\right)^{9/4}\,{\rm g}.\label{mass12}
 \end{equation}
 If $B_{\rm s}=2.2\times 10^{14}\,{\rm G}$, $M=1.4M_\odot$, and $R_*=10^6\,$cm are adopted, we obtain the magnetic interaction radius
 \begin{equation}
 R_{\rm m}\sim 1.2\times 10^7\left(\frac{\Delta t}{0.6\,{\rm ms}}\right)^{-1/18}\,{\rm cm},\label{Rm2}
 \end{equation}
 the cylindrical radius
 \begin{equation}
 r(R_{\rm m})\sim 2.1\times 10^4\left(\frac{\Delta t}{0.6\,{\rm ms}}\right)^{17/36}\,{\rm cm}, \label{rm2}
 \end{equation}
 and the inclination angle of an FRB-emitting region
 \begin{equation}
 \theta_{\rm i}\sim 1.7\times 10^{-3}\left(\frac{\Delta t}{0.6\,{\rm ms}}\right)^{19/36}.\label{theta2}
 \end{equation}
 Equations (\ref{Rm2})-(\ref{theta2}) give the parameters of the geometry of FRB 200428's emitting region in our model.

 Second, as for the radio properties, FRB 200428 was detected by the STARE2 telescope \citep{boch20}, implying that $\nu_{\rm curv}\sim 1.4\,$GHz, that is,
 \begin{equation}
 \gamma\sim 140\chi\left(\frac{\Delta t}{0.6\,{\rm ms}}\right)^{-5/72},\label{gamma2}
 \end{equation}
 and
 \begin{equation}
 \chi^3\delta\sim 0.6\left(\frac{\Delta t}{0.6\,{\rm ms}}\right)^{11/72}.\label{delta2}
 \end{equation}
 The isotropic-equivalent energy release becomes
 \begin{equation}
 E_{\rm radio} \sim 1.4\times 10^{35}\frac{\delta^3}{\chi^2}\left(\frac{\Delta t}{0.6\,{\rm ms}}\right)^{119/36}\,{\rm erg}.\label{Eradio2}
 \end{equation}
 A requirement of $E_{\rm radio}\sim 1.7\times 10^{35}(\Delta t/0.6\,{\rm ms})\,$erg leads to
 \begin{equation}
 \frac{\delta^3}{\chi^2}\sim 1.2\left(\frac{\Delta t}{0.6\,{\rm ms}}\right)^{-83/36}.\label{chi2}
 \end{equation}
 The solution of Equations (\ref{delta2}) and (\ref{chi2}) is
 \begin{equation}
 \chi\sim 0.85\left(\frac{\Delta t}{0.6\,{\rm ms}}\right)^{199/792},\label{chi3}
 \end{equation}
 and
 \begin{equation}
 \delta\sim1.0\left(\frac{\Delta t}{0.6\,{\rm ms}}\right)^{-119/198}.\label{delta3}
 \end{equation}
It can be seen that $\chi\lesssim 1$ and $\delta\sim 1$, showing that our model is self-consistent. This also implies that our line of sight is just within the solid angle of FRB 200428. In addition, combining Equations (\ref{gamma2}) and (\ref{chi3}), we find the electrons' typical Lorentz factor $\gamma\sim 120$ for $\Delta t\sim0.6\,$ms.

Third, as for the XRB properties, Equation (\ref{Eg}) shows
\begin{equation}
E_{\rm X}\sim 1.9\times 10^{40}\zeta\left(\frac{m_{\rm a}}{10^{20}\,{\rm g}}\right)\,{\rm erg},\label{Eg2}
\end{equation}
where $\zeta$ is the X-ray radiation efficiency and its upper limit is $\sim1/2$ because at least a half of the gravitational energy release $E_{\rm G}$ is transferred inward to the thermal energy of the stellar matter and eventually emitted by neutrinos. Equation (\ref{Eg2}) is consistent with the total energy of the observed XRB from SGR 1935+2154 as long as $\zeta(m_{\rm a}/10^{20}\,{\rm g})\sim 0.5$, indicating that our model can also well explain the XRB. For a more massive asteroid (i.e., $m_{\rm a}>0.5\times10^{20}\zeta^{-1}\,$g), this conclusion is more viable. On the other hand, as the asteroid collides with the stellar surface, a resultant hot fireball has such a high temperature $T_{\rm fb}$ estimated by Equation (\ref{Tfb}) that a dense population of e$^\pm$-pairs are inevitably created. Thermal X-rays from a photosphere in the relativistically expanding fireball are emitted and superposition of radiation from different photospheres could generate a multi-temperature blackbody spectrum of an XRB. In addition, collisions between different shells in the fireball may lead to nonthermal emission. It would be expected that these processes can explain the observed spectrum of the XRB \citep{Li2020,tav20,mere20,rid20}, as discussed in section \ref{model}. A full discussion of the spectrum is well beyond the scope of this Letter and will be left elsewhere.

Fourth, the above constraints are given for the first pulse of FRB 200428. For {\em the second pulse} of this burst, $\Delta t\sim 0.34\,$ms and $E_{\rm radio}\simeq [1/(1+\xi)]\times(E_{\rm CHIME}+E_{\rm STARE2})\sim 0.8\times 10^{35}(\Delta t/0.34\,{\rm ms})\,$erg for the distance $D\sim10\,$kpc. These observed data have been used to provide similar constraints on model parameters.

Fifth, the observed light curve of FRB 200428 requires that the time interval ($\sim28.9$\,ms) between two sub-millisecond pulses should be smaller than the total asteroidal accretion timescale $t_{\rm a}$ (i.e., Equation \ref{ttot}), implying that $m_{\rm a}>3\times 10^{18}\,$g. This constraint is naturally satisfied for the mass limit from XRB observations.

Finally, during an active period of 29 XRBs from SGR 1935+2154 observed by Fermi/GBM prior to FRB 200428, the FAST radio telescope observed the magnetar but did not detect any FRB \citep{Lin2020}. This non-detection result can be understood in our model: $\delta\simeq 1/(2\gamma^2)\sim 3.5\times 10^{-5}$ for $\theta_{\rm v}\gg \theta_{\rm i}$, in which case the isotropic-equivalent radio emission energy observed at $\theta_{\rm v}$ is $\sim0.7\times 10^{22}\,$erg even if a fragment has a similar mass. This energy is too low for the FAST telescope to be able to detect any FRB. Furthermore, for a less massive fragment, any FRB-like signal from SGR 1935+2154 at large $\theta_{\rm v}$ cannot be observed because of a lower intrinsical isotropic-equivalent energy release $E_{\rm radio}$.

\section{Conclusions}\label{con}
In this Letter, we have proposed a new model for the association of FRB 200428 with an XRB from SGR 1935+2154, in which a magnetar encounters an asteroid with mass of $\sim10^{20}\,$g. We have shown that such an impact can self-consistently interpret the emission properties of FRB 200428 and its associated XRB. This model is different from that of \cite{Dai2016}, because we here considered the magnetic interaction radius $R_{\rm m}$, at which the asteroid during its free infall must be impeded by an ultra-strong magnetic field and then accreted onto the poles along magnetic field lines, colliding with the stellar surface and generating an XRB. Although it is undetected at cosmological distances, such an XRB in the Galaxy is bright enough to be observed by current X-ray satellites, as discussed in \cite{Dai2016}. We constrained the model parameters. Our conclusions are summarized as follows.
\begin{itemize}
\item FRB 200428-emitting region looks like an openmouthed clam, whose inclination angle and magnetic interaction radius are $\theta_{\rm i}\sim 1.7\times 10^{-3}$ and $R_{\rm m}\sim 1.2\times 10^7\,$cm, respectively. The FRB emits along magnetic field lines around $R_{\rm m}$.
\item The typical Lorentz factor $\gamma\sim120$ of emitting electrons is found to understand a low isotropic-equivalent energy of FRB 200428 as compared to cosmological FRBs. Our line of sight is just within the solid angle of this burst. If the viewing angle is much larger than $\theta_{\rm i}$ (i.e., an off-plane case), the isotropic-equivalent energy release becomes extremely low. This is why the FAST telescope has not detected any FRB-like signal during the active phase of 29 XRBs observed by Fermi/GBM.
\item As the asteroid collides with the stellar surface, a resultant hot fireball has a temperature $\sim 1.8\times 10^{10}\,$K. This leads to a dense population of e$^\pm$ pairs. Superposition of thermal emission from different photospheres and nonthermal emission from collisions between different shells in the fireball is expected to account for the observed XRB's spectrum\footnote{After the submission of this Letter, \cite{Younes2020} suggested a possible polar origin of the FRB 200428-associated XRB, based on the fact that such an XRB has an extremely low occurrence rate (at most around 1 in 7000) and an unusual spectrum (e.g., the spectral cutoff energy is much higher than that of 24 XRBs emitted in 13 hours prior to the FRB). This suggestion is clearly consistent with the physical picture shown in Figure \ref{fig1}.}. In addition, from Equation (\ref{ttot}), the typical duration ($t_{\rm a}$) of an XRB is of order $\sim 0.1(m_{\rm a}/10^{20}\,{\rm g})^{1/3}\,$s, which is basically consistent with the X-ray observations.
\end{itemize}

What should be pointed out is that in this Letter we have only discussed the asteroid-neutron star direct collision case in which the impact area $\pi b^2$ (where $b$ is the so-called impact parameter) is smaller than the capture cross-section given by Equation (18) of \cite{Dai2016}, that is, $\pi b^2<\sigma_{\rm a}=2.4\times 10^{19}R_{*,6}(M/1.4M_\odot)V_{\rm p,7}^{-2}\,{\rm cm}^2$, where $R_{*,6}=R_*/10^6\,{\rm cm}$ and $V_{\rm p,7}=V_{\rm p}/10^7\,{\rm cm}\,{\rm s}^{-1}$, implying that $b<b_{\rm cr}=(\sigma_{\rm a}/\pi)^{1/2}=2.8\times 10^9R_{*,6}^{1/2}(M/1.4M_\odot)^{1/2}V_{\rm p,7}^{-1}\,{\rm cm}$. In this case, an asteroid is captured by a neutron star and then eventually impacts the stellar surface. If $b>b_{\rm cr}$, on the other hand, an asteroid cannot be captured and instead it flies out nearby the neutron star, in which case the asteroid still interacts with the stellar magnetosphere, perhaps leading to a faint burst-like electromagnetic signal due to the effect of a weak outer magnetic field.

\acknowledgments
I would like to thank the anonymous referee for his/her helpful comments that have allowed me to improve the presentation of this Letter. I also thank Lin Lin, Xiangyu Wang, Xuefeng Wu, Yunwei Yu, Bing Zhang, and Shuangnan Zhang for their useful discussions. This work was supported by the National Key Research and Development Program of China (grant No. 2017YFA0402600) and the National Natural Science Foundation of China (grant No. 11833003).

\end{document}